\documentclass[aps,prl,preprint,superscriptaddress]{revtex4}
\usepackage{graphicx}
\usepackage{xcolor}
\input epsf
\usepackage{amsmath,amssymb}
\begin{document}
\title{Stochastic resonance in thermally bistable Josephson weak-links and $\mu$-SQUIDs}
\author{Sagar Paul}
\affiliation{Department of Physics, Indian Institute of Technology Kanpur, Kanpur 208016, India}
\author{Ganesh Kotagiri}
\affiliation{Department of Physics, Indian Institute of Technology Kanpur, Kanpur 208016, India}
\author{Rini Ganguly}
\affiliation{\mbox{Univ.} Grenoble Alpes, CNRS, Grenoble INP, Institut N\'eel, 38000 Grenoble, France}
\author{Herv\'{e} Courtois}
\affiliation{\mbox{Univ.} Grenoble Alpes, CNRS, Grenoble INP, Institut N\'eel, 38000 Grenoble, France}
\author{Clemens B. Winkelmann}
\affiliation{\mbox{Univ.} Grenoble Alpes, CNRS, Grenoble INP, Institut N\'eel, 38000 Grenoble, France}
\author{Anjan K. Gupta}
\affiliation{Department of Physics, Indian Institute of Technology Kanpur, Kanpur 208016, India}
\date{\today}

\begin{abstract}
Constriction-based Josephson weak-links display a thermal bistability between two states exhibiting zero and finite voltages. This manifests in experiments either as hysteresis in weak-link's current voltage characteristics or as random telegraphic signal in voltage. In the latter case, a noise-driven amplification of a sinusoidal excitation of the device is observed, at frequencies matching the characteristic switching frequency in telegraphic signal, a phenomenon known as stochastic resonance. The observed behavior is understood using a two-state model of stochastic resonance and is exploited to illustrate an enhanced signal-to-noise-ratio in a $\mu$-SQUID as a magnetic field sensor.
\end{abstract}

\maketitle
\section{Introduction}

Periodically driven bistable systems exhibit an enhanced response when the noise induced stochastic transition rate between the two states of the bistable system matches with the drive frequency. This is the phenomenon of stochastic resonance (SR) that was first invoked as a plausible explanation for the periodic occurrence of `ice ages' \cite{Benzi_1981,wies-moss-persp}. It has since been studied in several bistable systems including electronic circuits \cite{fauve-heslot-ckt}, mesoscopic electronic systems \cite{vardi-double-dot,wagner-nature}, lasers \cite{mcnamara-ring-laser}, semiconductor devices \cite{mantegna-tunn-diode}, nano-mechanical oscillators \cite{bradzey-nature}, particle traps \cite{simon-optical-trap}, chemical systems \cite{hohman-SR-chem} as well as in superconducting quantum interference devices (SQUIDs) having bistability with respect to magnetic flux \cite{rf-SQUID-hibbs,SQUID-rouse}.

The random switching of a bistable system between two states leads to a random telegraphic noise (RTN) in a suitable measured quantity. The power spectral density of this RTN has a Lorenzian shape \cite{RTN-PSD-machlup} with a cutoff frequency $f_{\rm c}$ determined by the Kramer's rate \cite{Kramers}. The latter is characteristic of the switching between the two states. It depends both on the noise intensity and the energy barrier height between the two stable states. The phenomenon of stochastic resonance consists of a peak in the response to a periodic drive, as a function of the noise controlled cutoff frequency $f_{\rm c}$, when it matches the drive frequency. Thus there exists an optimum noise level that enhances the system response to an excitation \cite{sudeshna-prl}.

Josephson weak links (WL) based on a constriction between two superconductors have been of interest for their physics \cite{likharev-rmp} and applications such as radiation detectors \cite{rad-det} and nanoscale magnetometry \cite{SQUID-appl} using micron-size SQUID, {\mbox{i.e.}} $\mu$-SQUID \cite{SQUID-appl,Nikhil-prl}. Thus, a better understanding of these WLs is essential for further improvements in such devices. Constriction-based superconducting weak-links exhibit a dynamic thermal bistability \cite{DTM-JAP,sourav-PRB-dyn}, in addition to the static thermal bistabilities \cite{Nikhil-prl,Scocpol}. The former, of interest here, exists in certain bias current $I_{\rm b}$ range, starting from the dynamic re-trapping current $I_{\rm r}^{\rm dyn}$ and up to the critical current $I_{\rm c}$. The two possible WL states in this current range are the superconducting, zero-voltage state and the dissipative, non-zero voltage state. Well above a certain threshold temperature $T_{\rm h}$, a rapid switching between these two states leads to the disappearance of hysteresis and observation of an average intermediate voltage. In a $\mu$-SQUID, with two WLs in parallel forming a micron-scale loop, this enables the detection of a highly flux-sensitive voltage for magnetometry operation. This voltage-readout mode \cite{sourav-PRB-dyn} has advantages in terms of speed and sensitivity as compared to the more commonly used critical current measurement in a current-sweep mode \cite{SQUID-appl} in hysteretic devices. As the temperature is lowered, the switching frequency decreases, leading to an observable RTN in the WL voltage \cite{sourav-RTN}, before eventually entering the fully hysteretic state well below $T_{\rm h}$.

In this paper, we explore the transition of thermally bistable superconducting weak links from a  reversible regime towards a hysteretic regime. We observe random switching between two voltage states with a Lorenzian frequency spectrum whose cutoff frequency increases with temperature. AC driving at frequencies up to this cutoff frequency leads to a relatively large magnitude deterministic signal and with an enhanced signal to noise ratio. These observations are quantitatively understood using a two-state model \cite{mcnamara-ring-laser} of stochastic resonance. This enables us to demonstrate an enhanced signal-to-noise-ratio in a $\mu$-SQUID magnetic field sensor in a certain operating regime.

\section{Bi-stability, random telegraphic noise and stochastic resonance}

\begin{figure}[t!]
\centering
\includegraphics[width=0.9\columnwidth]{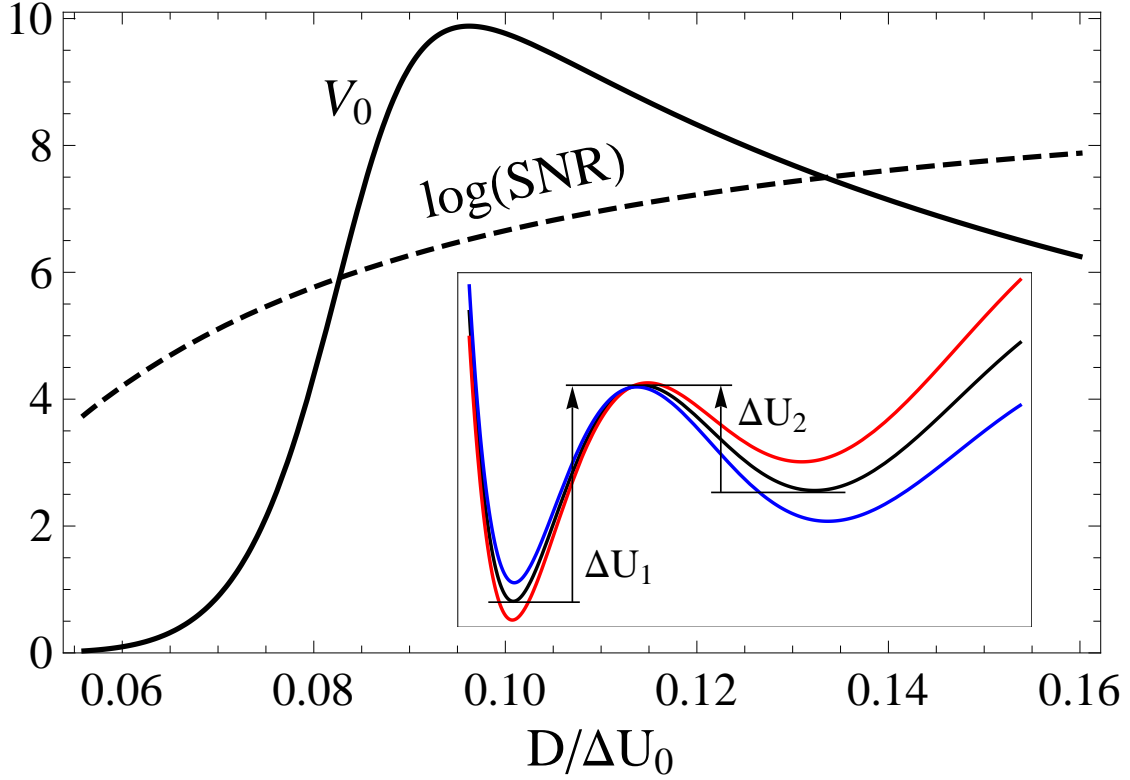}
\caption{Schematic variation of response $V_0$ to an AC current modulation and corresponding log(SNR) as a function of noise intensity $D$. The continuous line shows $(\Delta U_{\rm 0}/D)[1+(f_{\rm dr}/f_{\rm c})^2]^{-1/2}$ for $V_{\rm 0}$ and dashed line shows $\log_{10}[(\Delta U_{\rm 0}/D)^{2}f_{\rm c}]$ for log(SNR) with $f_{\rm c}=f_0\exp(-\Delta U_{\rm 0}/D)$. Here, $f_{\rm 0}=10^7$ Hz and $f_{\rm dr}=100$ Hz. The inset shows a schematic of the energy corresponding to the bias currents: $I_{\rm b0}$ (black), $I_{\rm b0}-\delta I$ (red), $I_{\rm b0}+\delta I$ (blue). $\Delta U_{\rm 1,2}$, marked for the black curve corresponding to the symmetric case for a given $D$, define the two barrier energies.}
\label{fig:two-state}
\end{figure}

Random telegraphic noise appears in systems exhibiting two (meta-) stable states $i=1,2$, separated by an energy barrier, as depicted in \mbox{Fig. \ref{fig:two-state}} inset, when exposed to a white noise of intensity $D$. The latter can arise from thermal fluctuations or any other noise source. The system will switch randomly from state-$i$ to the other at the classical Kramer's rate \cite{Kramers}, $r_{\rm i}=(\omega_{\rm i}\,\omega_{\rm b}/2\pi\gamma)\exp(-\Delta U_{\rm i}/D)$. Here, $\Delta U_{\rm i}$ is the barrier height seen from state-$i$ minimum, $\gamma$ is the viscous friction and $\omega_{\rm i}^2$, $\omega_{\rm b}^2$ are proportional to the second derivative of the potential at the corresponding potential minimum and at the peak of the barrier in-between two states, respectively. Thus for a fixed $\Delta U_{\rm i}$, the rate $r_{\rm i}$  will increase sharply with the noise $D$.

The system spends an average residency time $\tau_{\rm i}\sim 1/r_{\rm i}$ in the state $i$. This leads to probabilities $p_{\rm i}=\tau_{\rm i}/(\tau_1+\tau_2)$ of finding the system in the state $i$. For the WL system under study, the instantaneous voltage $V$ in state-1 is zero and in state-2 it is $v$ \cite{footnote-1}. Once averaged over a time larger than both $\tau_{\rm i}$'s, it is $V_{\rm av}=p_2 v$. The standard deviation of the measured $V$ will be given by $\sigma_{\rm V}^2=(1-p_{\rm 2})V_{\rm av}^2+p_{\rm 2}(V_{\rm av}-v)^2$, which on eliminating $p_{\rm 2}$ in favor of $v$ and $V_{\rm av}$ gives:
\begin{align}
\sigma_V^2=V_{\rm av}(v-V_{\rm av}).
\label{eq:eq1}
\end{align}
As shown in \mbox{Fig. \ref{fig:RTN}b}, this quadratic dependence is accurately observed in the experiments and allows for a precise determination of the symmetric point, defined by $\tau_{\rm 1}=\tau_{\rm 2}=\tau_{\rm 0}$, as the position of the maximum of $\sigma_{\rm V}^2$.

The spectral noise density \cite{psd-suppl} of the RTN in $V$ works out to be \cite{RTN-PSD-machlup} $S_{\rm V}^{\rm N}(f)=4v^2/[(\tau_{\rm 1}+\tau_{\rm 2})\{(\tau_{\rm 1}^{-1}+\tau_{\rm 2}^{-1})^2+(2\pi f)^2\}]$. Here, a delta-function peak at $f=0$, corresponding to non-zero $V_{\rm av}$, has been omitted. At the symmetric point we obtain \cite{suppl-info}:
\begin{align}
S_{\rm V}^{\rm N}(f)=\frac{v^2 f_{\rm c}}{2\pi(f^2+f_{\rm c}^2)}
\label{eq:eq2}
\end{align}
with the cutoff frequency $f_{\rm c}=1/\pi\tau_{\rm 0}$. The total noise power, integrated over all frequencies, matches with the variance $\sigma_{\rm V}^2=v^2/4$ of the symmetric case for which $V_{\rm av}=v/2$, see \mbox{Eq. (\ref{eq:eq1}}).

We elaborate further on stochastic resonance from the superconducting WL viewpoint using the two-state model of McNamara {\it et al.} \cite{mcnamara-ring-laser,SR-rev-RMP}. In a WL, the drive parameter is the bias current $I_{\rm b}$. When the latter deviates by a small amount $\delta I$ from the symmetric value $I_{\rm b0}$ defined by $\tau_{\rm 1}=\tau_{\rm 2}=\tau_{\rm 0}$ (see \mbox{Fig. \ref{fig:two-state} inset}), this induces an imbalance of the $\tau_{\rm i}$'s which we write up to linear order in $\delta I$ as $\tau_{\rm i}=\tau_{\rm 0}(1\mp\alpha_{\rm i}\delta I)$. As the Kramer's rate dependence on $I_{\rm b}$ is predominantly determined by that of $\Delta U_{\rm i}$, as compared to that of $\omega_{\rm i}$, we write:
\begin{align}
\alpha_{\rm i} \approx \mp\frac{1}{D}\left[\frac{\partial\Delta U_{\rm i}}{\partial I_{\rm b}}\right]_{I_{\rm b}=I_{\rm b0}}.
\label{eq:eq2a}
\end{align}
For a bias current $I_{\rm b}$ close to the symmetric point, we get $V_{\rm av}=v\tau_{\rm 2}/(\tau_{\rm 1}+\tau_{\rm 2})=(v/2)[1+ \alpha\delta I]$ with $\alpha=(\alpha_{\rm 1}+\alpha_{\rm 2})/2$. The slope of the DC current-voltage characteristics (IVC) at $I_{\rm b0}$ is thus given by $dV_{\rm av}/dI=\alpha v/2$. This parameter $\alpha$, with $v$ known from the peak value of $\sigma_V^2$, can thus be determined experimentally from a (slowly acquired) IVC.

We now consider a periodic drive $\delta I=\delta I_{\rm 0}\cos (2\pi  f_{\rm dr} t)$ of the bias current about $I_{\rm b0}$, which creates a periodic modulation in $\tau_{i}$ at the drive frequency $f_{\rm dr}$. We assume the drive to be small enough to perturb the potential only minutely so as not to cause any transitions by itself. We also consider the adiabatic limit: the drive frequency is kept much smaller than the intrinsic frequencies $\omega_{\rm 1,2}$, so that no transition can occur by way of deterministic dynamics. The ensemble-averaged response in the WL voltage works out \cite{suppl-info,SR-rev-RMP} as $\langle V(t)\rangle=\frac{v}{2}+V_{\rm 0}\cos(2\pi f_{\rm dr}t-\theta)$ with the amplitude $V_{\rm 0}$ given by:
\begin{align}
V_{\rm 0}=\delta I_0 \frac{\alpha v}{2}\sqrt{\frac{f_{\rm c}^2}{f_{\rm dr}^2+f_{\rm c}^2}}.
\label{eq:eq4}
\end{align}
Here, the phase $\theta$ follows $\tan\theta=f_{\rm dr} / f_{\rm c}$. The maximum response occurs at small frequencies $f_{\rm dr} \ll f_{\rm c}$ with a magnitude $V_{\rm 0}(f_{\rm c},0)=\delta I_{\rm 0} (\alpha v/2)$, \mbox{i.e.} the slope of the IVC times the current modulation amplitude. The voltage response is similar to a second order low pass filter with a bandwidth limited by the cut-off frequency $f_{\rm c}$. The overall power spectral density in $V$ is \cite{SR-rev-RMP}:
\begin{align}
S_{\rm V}(f)= [1-(2V_0^2/v^2)]S_{\rm V}^{\rm N}(f)+(V_0^2/2)\delta(f-f_{\rm dr})
\label{eq:eq40}.
\end{align}
Compared to the equilibrium state with no excitation, a fraction $2V_{\rm 0}^2/v^2$ of the noise power gets transformed into a signal power at $f_{\rm dr}$, so that the total power still remains $v^2/4$. The signal to noise ratio (SNR), defined as the ratio of the signal power to the noise power in a 1 Hz bandwidth at the signal frequency, works out as:
\begin{align}
\text{SNR} =\frac{\pi}{4}(\alpha\delta I_0)^2 f_{\rm c}.
\label{eq:eq6}
\end{align}

The behavior of both the response amplitude $V_{\rm 0}$ and the SNR can be discussed qualitatively with simple assumptions. The cutoff frequency $f_{\rm c}$ increases strongly with the noise amplitude $D$, with a dependence in between $\exp(-\Delta U_{\rm 1}/D)$ and $\exp(-\Delta U_{\rm 2}/D)$ \cite{comment-DeltaU}, which we assume to be $\exp(-\Delta U_{\rm 0}/D)$ with $\Delta U_{\rm 0}$ in between $\Delta U_{\rm 1}$ and $\Delta U_{\rm 2}$. At the same time, the parameter $\alpha$ decreases with $D$, but only as $D^{-1}$, see \mbox{Eq. (\ref{eq:eq2a})}. Overall, we find SNR $\propto D^{-2}f_{\rm c}(D)$. The SNR thus {\em increases} with the noise intensity $D$. This occurs in spite of the signal magnitude reduction as the spectral-density of RTN reduces faster with $D$ than the signal. Practically, the experiments will also have noise from other sources, such as measurement electronics, and at some point the noise from other sources will dominate over that of RTN. As a result, the actual measured SNR will decrease after this point.

From \mbox{Eqs. (\ref{eq:eq2a},\ref{eq:eq4})}, we find $V_{\rm 0}\propto D^{-1}[1+\{f_{\rm dr}/f_{\rm c}(D)\}^2]^{-1/2}$. The behavior of $V_{\rm 0}$ and SNR as a function of the noise amplitude $D/\Delta U_{\rm 0}$ and at a fixed drive frequency $f_{\rm dr}$ is displayed in \mbox{Fig. \ref{fig:two-state}}. A peak in $V_{\rm 0}$ is observed when the cut-off frequency matches the drive frequency, \mbox{i.e.} $f_{\rm c}=f_{\rm dr}$. A more quantitative analysis requires the detailed knowledge of the whole potential profile, including $\Delta U_{\rm i}$, $\omega_i$, $\omega_b$ and $\gamma$, together with its dependence on the current bias $I_{\rm b}$.

\section{Thermal bistability in weak links}

Let us now discuss the physics of Josephson WLs while keeping the stochastic resonance in view. A Josephson WL \cite{likharev-rmp} consists of a short region concentrating the phase drop $\varphi$ between two bulk superconducting condensates. It can be formed by a tunnel-junction, a normal-metallic link or simply a narrow superconducting constriction. The IVC of such WL devices often show hysteresis. During current ramp-up, the voltage switches from zero to a non-zero value at the critical current $I_{\rm c}$. During current ramp-down from the finite-voltage state, hysteretic devices transit back to a zero voltage at a re-trapping current less than the critical current. In the present work, the bistability refers to the zero and non-zero voltage states of the WL whereas earlier stochastic resonance studies in SQUIDs \cite{rf-SQUID-hibbs,SQUID-rouse} pertain to the bistability with respect to magnetic flux.

In tunnel-junction based Josephson junctions, the resistively and capacitively shunted junction (RCSJ) model \cite{tinkham-book} successfully describes the transport characteristics and other aspects of the phase $\varphi$ dynamics. This model leads to a tilted-washboard potential $-(\hbar I_{\rm c}/2e)(\cos\varphi+I_{\rm b}/I_{\rm c})$ exhibiting minima (for $I_{\rm b}<I_{\rm c}$) at a $\Delta\varphi=2\pi$ interval and a tilt proportional to the current bias $I_{\rm b}$. The $\varphi$-dynamics is then described by the motion, in this potential, of a particle with position $\varphi$, mass $(\hbar/2e)^2C$, with $C$ as junction capacitance, subjected to a drag force proportional to the junction conductance. In the zero voltage state, $\varphi$ takes a static value at a potential-minimum, separated by energy barriers from neighboring minima. The height of the barrier, in-between two neighboring minima, reduces with increasing bias current $I_{\rm b}$ and vanishes at $I_{\rm b}=I_{\rm c}$. The barrier height for $I_{\rm b}$ close to $I_{\rm c}$ is given by ($\hbar I_{\rm c}/2e)(1-I_{\rm b}/I_{\rm c})^{3/2}$ \cite{tinkham-book}. Any transition to a neighboring minimum, through thermal activation or quantum tunneling, amounts to a change in $\varphi$ by $2\pi$ and is called a phase-slip event. Such an event leads to a voltage pulse and thus Joule heat deposition. In the non-zero voltage steady-state, $\varphi$ evolves continuously with time, leading to a voltage $V=(\hbar/2e)(d\varphi/dt)$. According to this model, hysteresis arises when the $C$-dependent inertial term in the equation of motion for $\varphi$ dominates over the drag term.

Constriction-type WLs have negligible capacitance $C$, or mass term in the RCSJ model, but their IVC often still show hysteresis. This is attributed to heat generated \cite{PRL-Courtois-2008} by the large critical current in such WLs, hence a large dissipated power $I_{\rm c}V$ at the voltage onset, together with the poor heat evacuation from such a nanoscale object. A dynamic thermal model (DTM), incorporating heating effects together with phase dynamics, describes well the bistability between the zero and finite voltage states that is found within a bias current range $I_{\rm r}^{\rm dyn}<I_{\rm b}<I_{\rm c}$ \cite{DTM-JAP}. The so-called dynamic re-trapping current $I_{\rm r}^{\rm dyn}$ never exceeds the critical current $I_{\rm c}$ but the two approach zero as the bath temperature $T$ approaches the critical temperature $T_{\rm c}$. The extent of the bistable region in bias current, \mbox{i.e.} $I_{\rm c}-I_{\rm r}^{\rm dyn}$, decreases with increasing temperature while both states become more susceptible to noise. As a result, the WL switches randomly between the two states above a crossover temperature $T_{\rm h}$, leading to a random telegraphic noise in voltage.

Although one can understand the bistability using the DTM, evaluating the susceptibility of the two states to noise requires a detailed understanding of the effective potential or free energy of the WL, which is coupled to a thermal bath, and as a function of bias current. The theoretical approaches on the switching statistics of the two states in Josephson weak links due to noise are based on the tilted washboard potential \cite{ben-jacob+valenti} that ignore heating effects. In fact, there is no such general formulation of free-energy for such a dissipative and open dynamical system. Qualitatively, any transition from the zero voltage state to the dissipative state is triggered by a phase-slip. However, every phase slip may not eventually lead to the steady dissipative state as it requires accumulation of more heat than that of a single phase-slip \cite{Shah-PRL}. Nevertheless, for the experimental results and analysis reported here, the detailed knowledge of the free-energy, though insightful, is not essential for the validity of the two-state model. All the required parameters, such as the temperature dependent $\alpha$ and $f_{\rm c}$, can be extracted from the IVC and noise power spectral density.

\section{Experimental details}

We studied superconducting WL devices made of two bulk-like superconductors connected by a short and narrow constriction as shown in the inset of \mbox{Fig. \ref{fig:IVC}(a)} with two current and two voltage pads that are more than 100 $\mu$m away from the WL. These were fabricated using e-beam evaporated 20-nm-thick films of Nb on a Si substrate by patterning an aluminum mask through electron beam lithography and liftoff \cite{sourav-PRB-dyn}. The aluminum pattern was then transferred to Nb by reactive ion etching, followed by wet aluminum removal. The patterned width of the WL is 30 nm. Four-probe electrical measurements were carried out in a 1.3 K base-temperature closed-cycle refrigerator. All the sample leads in this setup pass through a low temperature copper powder filter and room temperature pi-filter for reducing noise interference.

The measurement electronics consists of a ground-isolated current source for current bias and a commercial voltage amplifier with 100 kHz bandwidth for sample voltage amplification. The amplified voltage was digitized with electronics capable of 250 kS/s maximum sampling rate. A lock-in amplifier was used for the AC current bias and phase sensitive AC voltage detection. An anti-aliasing filter with low-pass cutoff of 50 kHz and 12 dB/Octave roll-off was used after the voltage amplifier. The voltage data for spectral analysis were acquired at 100 kS/s acquisition rate. The power spectral density of voltage $S_{\rm V}(f)$ was obtained from the voltage time series after mean subtraction. The reported $S_{\rm V}(f)$ is the spectral-power averaged over about one hundred time-series data. The voltage's mean and standard deviation were calculated from 5000-point voltage time series at 200 kS/s for measuring DC current-voltage characteristics of the WL. Here, we report our measurements and analysis on one WL device and a $\mu$-SQUID while we have observed the same behavior in five other WL devices. For $\mu$-SQUID measurements, the constant offset magnetic field was provided using a superconducting electromagnet built in the closed cycle refrigerator capable of 2 T magnetic field at 10 A current. A small low inductance superconducting coil, closely coupled to the device \cite{suppl-info}, was used to provide the oscillating magnetic field. The maximum frequency of the oscillating magnetic field was limited to $\sim$ 2 kHz due to eddy current heating of the metallic sample holder at higher frequencies.

\section{Time-averaged WL characteristics and RTN}

\begin{figure}[t!]
	\centering
	\includegraphics[width=0.96\columnwidth]{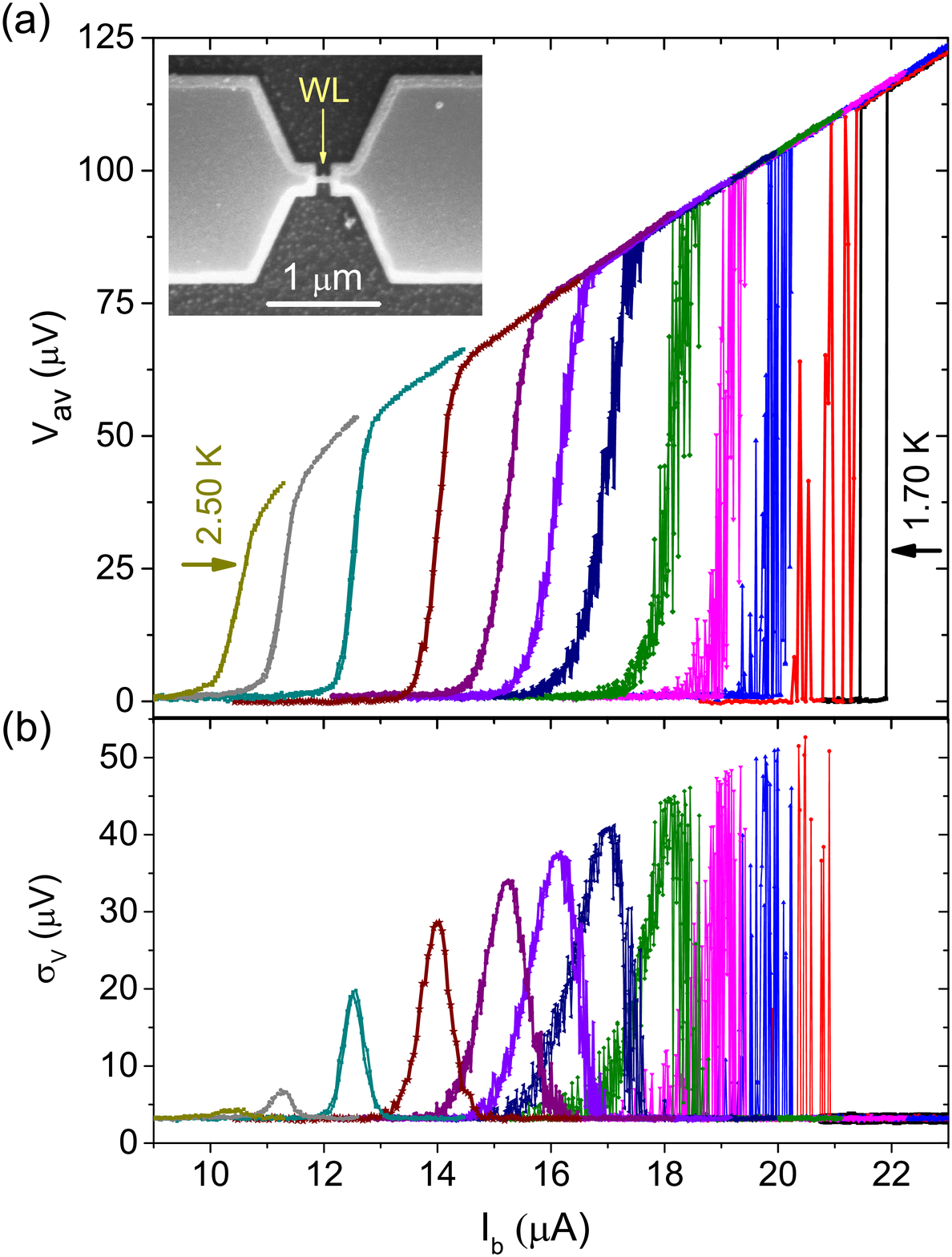}
	\caption{(a) $V_{\rm av}(I_{\rm b})$, {\mbox{i.e.}} IVCs, at bath temperatures $T$ = 2.50, 2.42, 2.31, 2.20, 2.10, 2.04, 1.99, 1.93, 1.86, 1.80, 1.75 and 1.70 K. The inset shows the scanning electron micrograph of the measured WL device. (b) Standard deviation in voltage $\sigma_{\rm V}$ as a function of the bias current $I_{\rm b}$, corresponding to the IVCs in (a). Each $V_{\rm av}$ and $\sigma_{\rm V}$ data point is found from 5000 voltage samples acquired at 200 kS/s rate.}
	\label{fig:IVC}
\end{figure}

Figure \ref{fig:IVC}(a) shows the IVCs in a current range encompassing $I_{\rm c}$ and $I_{\rm r}^{\rm dyn}$ of the studied Nb WL device for temperatures $T$ from 2.5 to 1.7 K. Note that both $I_{\rm c}$ and $I_{\rm r}^{\rm dyn}$ change by nearly a factor of two over this temperature range. In this temperature range, the experimental IVCs are seen to change from smooth and reversible to hysteretic behavior. On the high temperature side $T > T_{\rm h}$ (see below), random switchings occur in the current range $I_{\rm r}^{\rm dyn}<I_{\rm b}<I_{\rm c}$ between the zero and the finite voltage states of the WL. Here the mean interval between switches ($\sim\tau_{\rm i}$) is short compared to the voltage averaging time of 25 ms. A smooth IVC is thus observed without hysteresis, see data at 1.9 K and above. As the current bias $I_{\rm b}$ increases from $I_{\rm r}^{\rm dyn}$ to $I_{\rm c}$, the WL spends more time in the finite voltage state and thus $V_{\rm av}$ increases smoothly from zero to $v$.

\begin{figure}[t!]
	\centering
	\includegraphics[width=0.96\columnwidth]{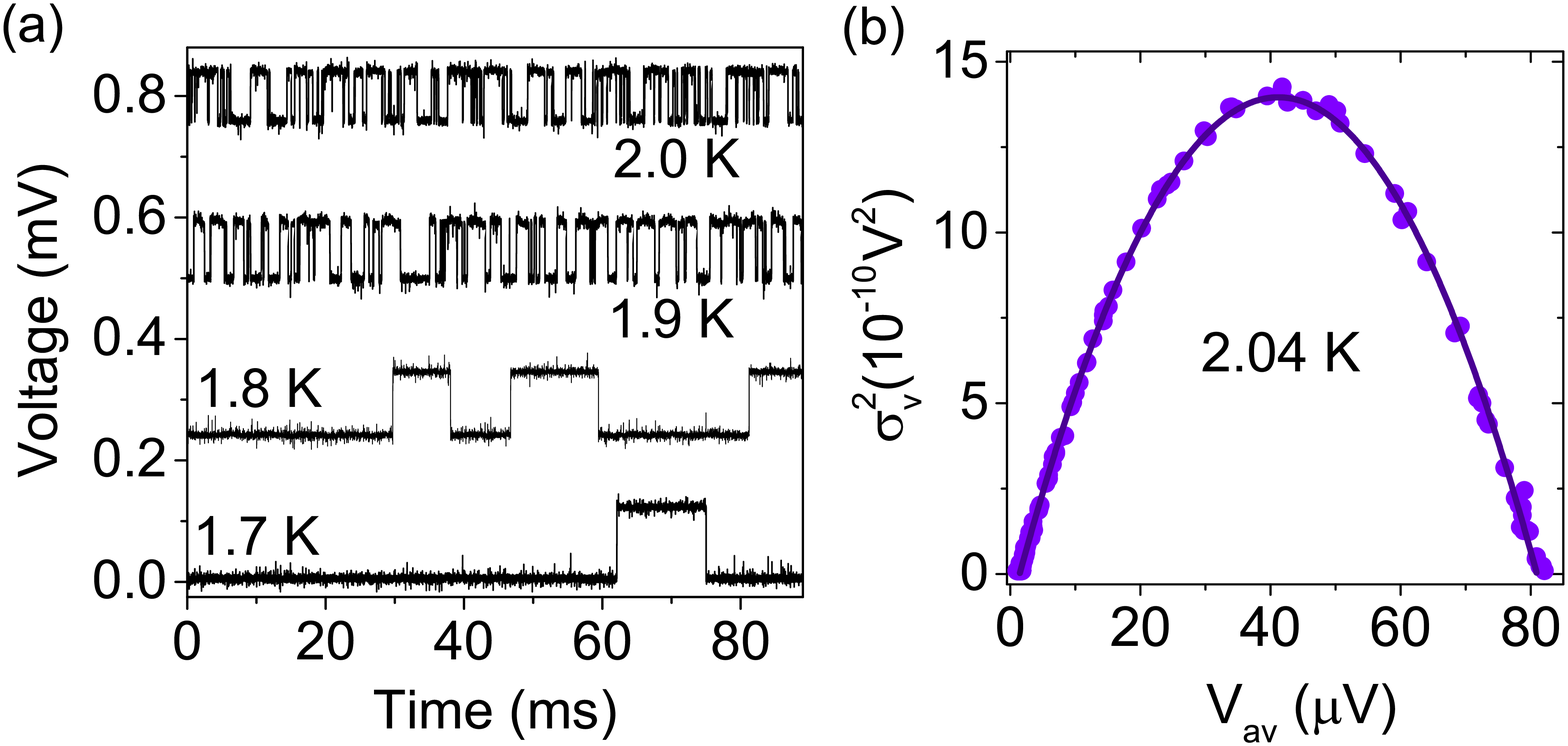}
	\caption{(a) Time traces of the voltage signal at the symmetric bias and at different temperatures plotted with equal offsets for clarity. (b) Variance $\sigma_{\rm V}^2$ versus average voltage $V_{\rm av}$ at 2.04 K with the dots as the experimental data and solid line as the fit to \mbox{Eq. (\ref{eq:eq1})}.}
	\label{fig:RTN}
\end{figure}

In a narrow temperature window near the threshold temperature $T_{\rm h} \sim 1.8$ K, the random switches between the zero- and finite-voltage states occur at a rate in-between the voltage sampling time and the averaging time. This can be seen in \mbox{Fig. \ref{fig:RTN}(a)} as RTN in the voltage time series taken at $I_{\rm b0}$. A time series consists of discrete voltage readings taken at 5 $\mu$s interval with each being the result of voltage sampling by the A/D card for about 2 $\mu$s which is significantly smaller than $\tau_{\rm i}$. Given the uncorrelated nature of the switchings, from the analysis presented further, a repeat of such time-series will give another realization of the set of switches. If the mean switching time is comparable to the voltage averaging time, a significant spread in $V_{\rm av}$ and $\sigma_{\rm V}$ will be obtained as evident in \mbox{Fig. \ref{fig:IVC}} for temperatures between 1.75 and 1.9 K. Such a regime of stochastic voltage response of a WL has been used to generate random numbers \cite{gambling-foltyn}.

At lower temperature $T < T_{\rm h}$ (see data at 1.7 K), a fully hysteretic IVC is observed as the switching time, again considered in the current range $I_{\rm r}^{\rm dyn}<I_{\rm b}<I_{\rm c}$, is larger than the voltage averaging time of 25 ms. As a consequence, the dissipative state is reached at $I_{\rm c}$ and the superconducting one recovered at $I_{\rm r}^{\rm dyn}$.

The criteria used for the estimation of $T_{\rm h}$ relies on the magnitude of voltage averaging time. A faster IVC measurement would have shown a transition to a hysteretic IVC at higher temperature. The definition of the threshold temperature $T_{\rm h}$ is therefore slightly measurement bandwidth dependent, in analogy to the blocking temperature in super-paramagnets \cite{super-paramag} at which a crossover from hysteretic to non-hysteretic behavior is seen.

The voltage standard deviation $\sigma_{\rm V}$ depends parabolically on the average voltage $V_{\rm av}$, as predicted by \mbox{Eq. (\ref{eq:eq1})}, see \mbox{Fig. \ref{fig:RTN}(b)}. The maximum of $\sigma_{\rm V}^2$ occurs at the current bias point $I_{\rm b}=I_{\rm b0}$, where the RTN is symmetric with $\tau_{\rm 1}=\tau_{\rm 2}$, $V_{\rm av}=v/2$ and $\sigma_{\rm V}^2=v^2/4$. All the noise and AC measurements were carried out at this symmetric bias current $I_{\rm b0}$ for ease of analysis. The two bias current $I_{\rm b}$ values at which $\sigma_{\rm V}$ (due to RTN) goes to zero correspond to $I_{\rm r}^{\rm dyn}$ and $I_{\rm c}$. With increasing bath temperature $T$, the critical current $I_{\rm c}$ reduces and thus both the voltage $v$ just above $I_{\rm c}$ and the variance $\sigma_{\rm V}^2$ decrease. For $T>2.2$ K, a faster decline in $\sigma_{\rm V}$ is observed, see \mbox{Fig. \ref{fig:IVC}(b)}, which comes from the switching rates $\tau_{\rm i}^{-1}$ exceeding the voltage measurement bandwidth.

\begin{figure}[t!]
\centering
\includegraphics[width=0.9\columnwidth]{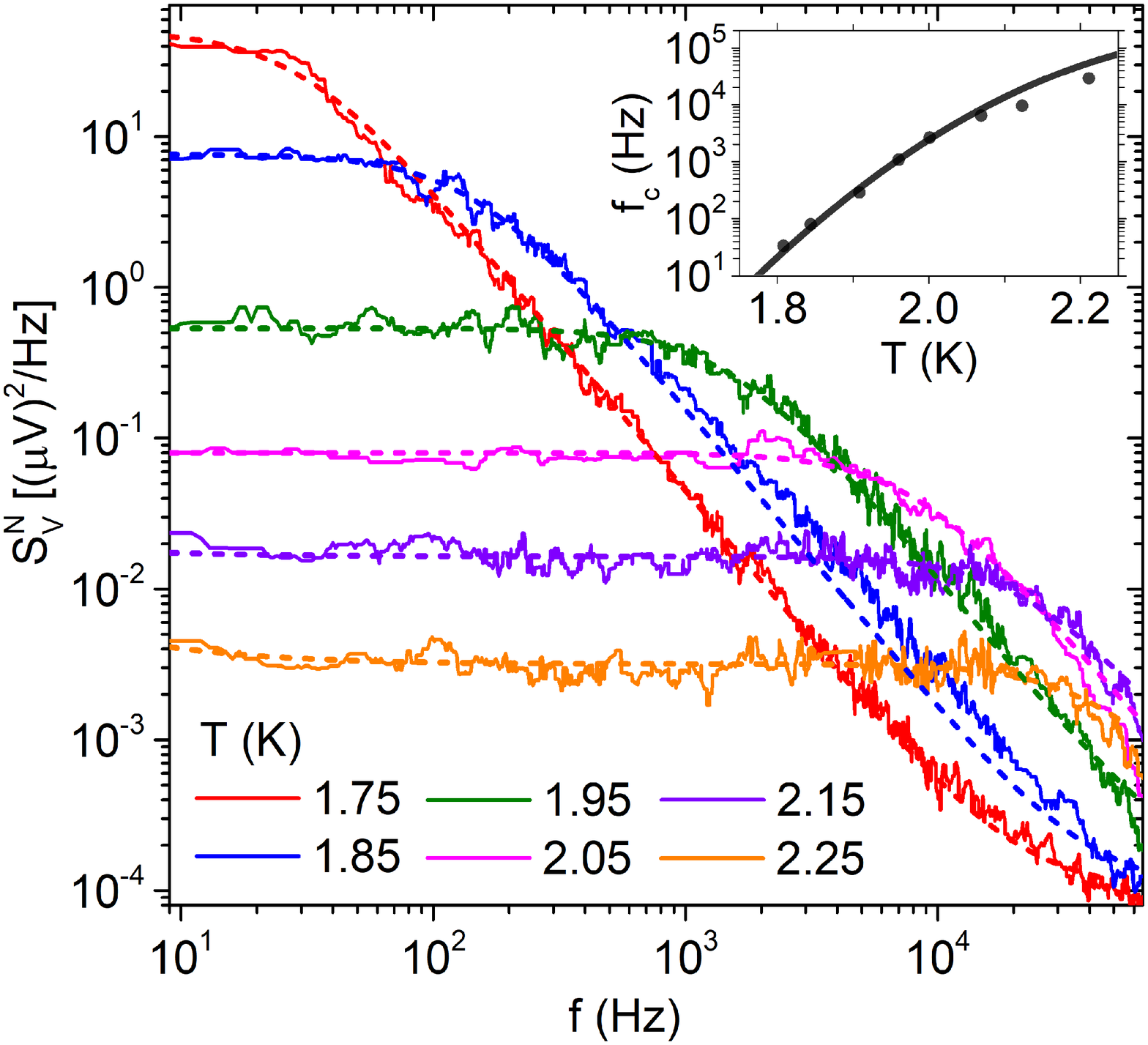}
\caption{Power spectral density $S_{\rm V}(f)$ of the WL voltage signal at the symmetric bias point and at different temperatures. The data are shown as full lines and the corresponding Lorenzian fits are displayed as dashed lines. The inset shows the variation of $f_{\rm c}$, obtained from the Lorenzian fits, with temperature. The data is shown with symbols and a parabolic fit by a continuous line.}
\label{fig:S-vs-f}
\end{figure}

\mbox{Figure \ref{fig:RTN}(a)} shows that the switching rates rapidly increase with increasing temperature. Consistently, the low frequency power spectral density $S_{\rm V}(f)$ decreases with increasing temperature, see \mbox{Fig. \ref{fig:S-vs-f}}. A Lorenzian analysis following \mbox{Eq. (\ref{eq:eq2})} was performed, including a white noise contribution of magnitude 12 nV/$\sqrt{Hz}$ and a $1/f$ noise. These extrinsic contributions arise from sources other than the WL, such as the voltage preamplifier. The $1/f$ component is negligible except for the highest temperatures, where it leads to a slight upturn at low frequencies. Besides these fixed extrinsic contributions, and because $v$ is known from the $\sigma_{\rm V}^2$ analysis, the cutoff frequency $f_{\rm c}$ is the only free parameter here. This, and the absence of any sharp peaks in measured $S_{\rm V}(f)$, clearly establishes the RTN behavior of the voltage. Interestingly, $f_{\rm c}$ is found to rise nearly exponentially with the bath temperature $T$, growing by four orders of magnitude in a rather narrow $\sim 0.5$ K temperature window, see \mbox{Fig. \ref{fig:S-vs-f}} inset.

\section{Response to a periodic current drive and stochastic resonance}

\begin{figure}[t!]
\centering
\includegraphics[width=0.95\columnwidth]{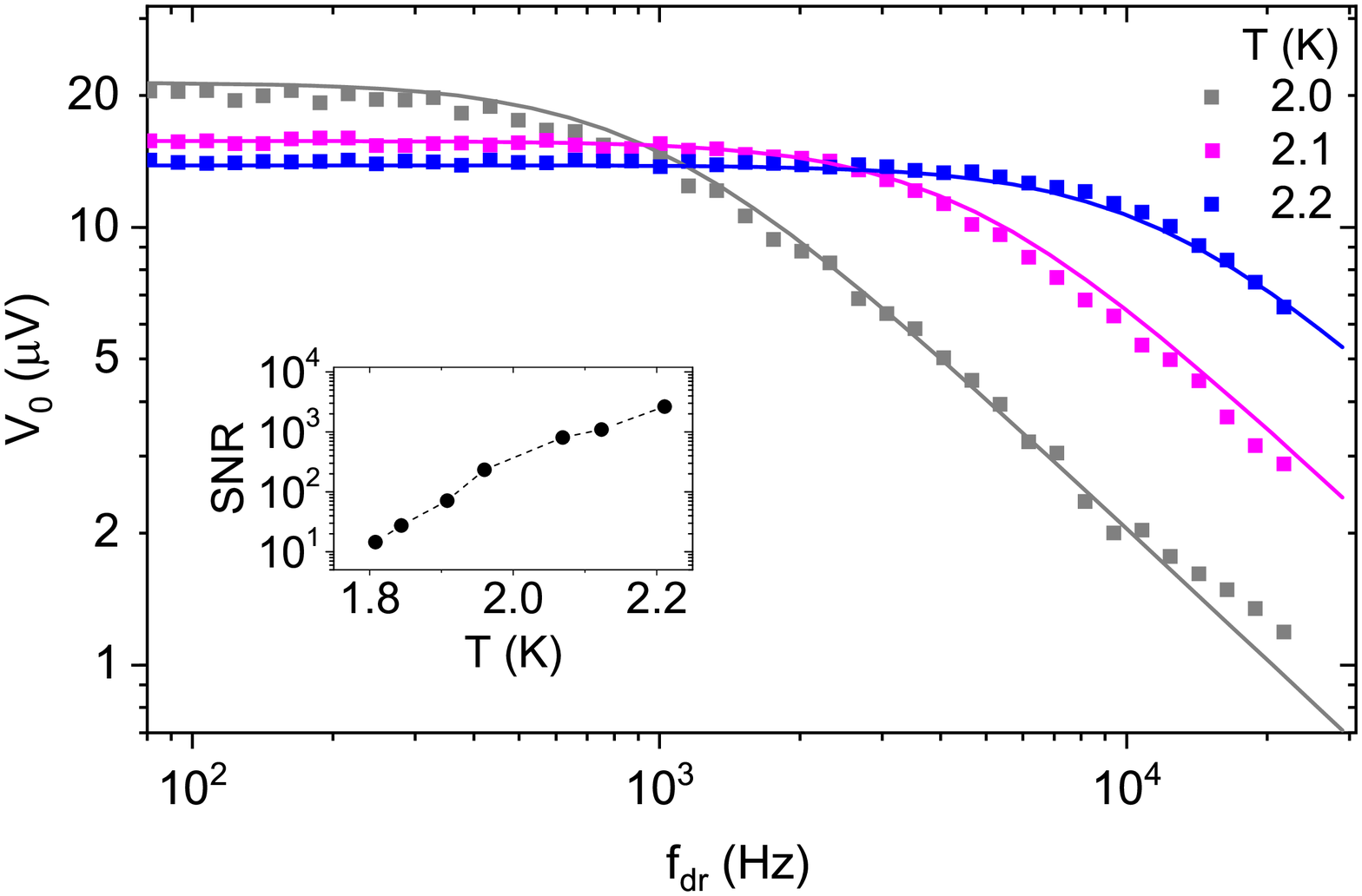}
\caption{Voltage response amplitude $V_{\rm 0}$ of the WL as a function of the bias drive frequency $f_{\rm dr}$ at different temperatures for a fixed AC drive amplitude $\delta I_{\rm 0}$ = 0.21 $\mu$A at the symmetric point. The symbols are the data and the solid lines are calculated using \mbox{Eq. (\ref{eq:eq4})}. Note the logarithmic scale for both the axes. The inset shows the measured variation of SNR (at $f_{\rm dr}$ = 100 Hz) with bath temperature.}
\label{fig:Vout(fdr)_1}
\end{figure}

For determining the device response to a periodic current drive, we add to the current bias $I_{\rm b0}$ at the symmetric point a small AC current at a frequency $f_{\rm dr}$ and with a constant amplitude  $\delta I_{\rm 0} =0.21\,\mu$A. The latter value is chosen to be small compared to the extent $I_{\rm c}-I_{\rm r}^{\rm dyn}$ of the bistable region, which secures the small perturbation limit of the stochastic resonance. The voltage response amplitude $V_{\rm 0}$ at the frequency $f_{\rm dr}$ was measured using a lock-in amplifier at different bath temperatures. The frequency response is found to follow the expected behavior of a second order low pass filter with a cutoff frequency $f_{\rm c}$, see \mbox{Fig. \ref{fig:Vout(fdr)_1}}. The solid lines in this plot show the prediction from \mbox{Eq. (\ref{eq:eq4})} with no fitting parameter, as both $\alpha v/2$ and $f_{\rm c}$ values at each bath temperature $T$ are known from the preceding analysis.

The SNR was experimentally determined from the power spectral density $S_{\rm V}^{\rm N}(f)$ by taking the ratio of the signal power to the noise power at a fixed drive frequency \cite{suppl-info}. \mbox{Figure \ref{fig:Vout(fdr)_1}} inset shows a nearly exponential rise in SNR with temperature. The relative rise in SNR, over the same 0.5 K temperature range, is less than that of $f_{\rm c}$ as seen in the inset of \mbox{Fig. \ref{fig:S-vs-f}}. This can be understood from the reduction in $\alpha$ with temperature, see \mbox{Eq. \ref{eq:eq6}}.

\begin{figure}[t!]
\centering
\includegraphics[width=0.9\columnwidth]{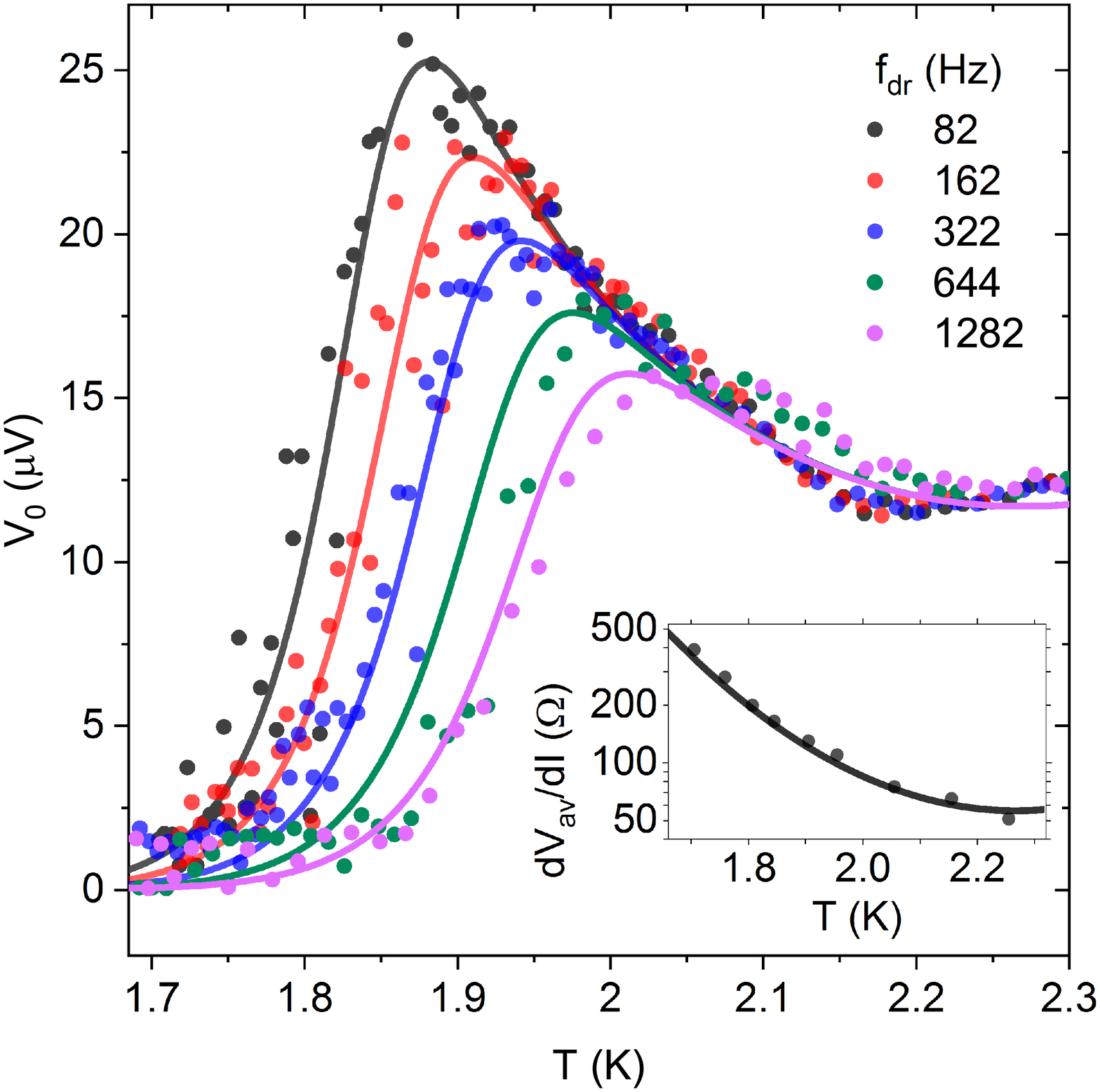}
\caption{Amplitude of the WL response $V_{\rm 0}$ to a periodic bias drive of amplitude 0.21 $\mu$A at the symmetric bias point as a function of the bath temperature for different drive frequency $f_{\rm dr}$ values. The symbols are the data. The inset shows in a semi-logarithmic scale the temperature dependence of the IVC slope, which is equal to $\alpha v/2$, together with a quadratic fit. This fit and that of $f_{\rm c}(T)$, shown in the inset of \mbox{Fig. \ref{fig:S-vs-f}}, are used as the sole parameters of the main panel calculated curves (continuous lines) using \mbox{Eq. (\ref{eq:eq4})}.}
\label{fig:Vout_T}
\end{figure}

In order to evidence the phenomenon of stochastic resonance, we show in \mbox{Fig. \ref{fig:Vout_T}} the AC response as a function of temperature and at several different drive frequency $f_{\rm dr}$ values. The fits of $f_{\rm c}(T)$ and $\alpha v/2(T)$ (solid lines in \mbox{Fig. \ref{fig:S-vs-f}} and \ref{fig:Vout_T} insets) were used in calculating the temperature dependent $V_{\rm 0}$, shown by continuous lines in \mbox{Fig. \ref{fig:Vout_T}} at the same $f_{\rm dr}$ values. The experimental data agree well with the calculations. From \mbox{Eq. (\ref{eq:eq4})}, we note that the initial sharp rise in $V_{\rm 0}$ with temperature comes from a nearly exponential increase in $f_{\rm c}$ while the decline beyond the peak occurs due to a reduction in $\alpha v/2$, \mbox{i.e.} the IVC-slope. Eventually the peak occurs close to a temperature at which $f_{\rm c}=f_{\rm dr}$. We note that the slope of the IVC just above $I_{\rm c}$ leads to a small variation in $v$ at the drive frequency. This slope is less than 5\% of the slope of the transition region below $I_{\rm c}$. We have ignored this small contribution to $V_{\rm 0}$ in the calculated curves which may be responsible for the non-zero $V_{\rm 0}$ at low temperatures.

This observed behavior is qualitatively similar to the expected response in the presence of a noise intensity $D$ depicted in \mbox{Fig. \ref{fig:two-state}}. For most studied experimental stochastic resonance systems \cite{fauve-heslot-ckt,mcnamara-ring-laser,hohman-SR-chem}, one varies noise ($D$ or $k_{\rm B} T$) keeping a constant $\Delta U$ in order to analyze the response at a fixed $f_{\rm dr}$ as a function of $D$. This provides the most direct evidence of stochastic resonance. In our case, the extrinsic noise of non-thermal origin is essentially constant, so that the noise amplitude is driven by temperature mainly. The thermal noise variation over a rather narrow bath temperature range combined with the decrease of the potential barrier with temperature is enough to probe the full width of the stochastic resonance peak over a range of drive frequencies. Moreover, the measured increase in $f_{\rm c}$ and decrease in $\alpha v/2$ quantitatively capture the measured temperature dependence of AC response as arising from stochastic resonance.

\section{Stochastic resonance with a periodic magnetic field drive in a $\mu$-SQUID}

In a $\mu$-SQUID consisting of two parallel WLs, the magnetic flux leads to a periodic modulation in the critical current with a flux-periodicity $\Phi_{\rm 0}$. This effectively modulates the scaled bias current $I_{\rm b}/I_{\rm c}$ that drives the SQUID behavior, as captured in the dynamic thermal model \cite{DTM-JAP}.

A periodic magnetic field drive was used, instead of a bias current drive, to measure the voltage response of a $\mu$-SQUID in its bistable region. A constant biasing magnetic field $B_{\rm 0}$ is first applied to submit the $\mu$-SQUID to a magnetic flux of a quarter of a flux quantum $\Phi_{\rm 0}$. In that case, the critical current response to the magnetic field is largest and linear. Moreover, the bias current is tuned to the symmetric point so that the IVC is also linear within a certain current range. Under these two combined conditions, the SQUID response $dV_{\rm av}/dB(T)$ can be written as $dV_{\rm av}/dI \times dI_{\rm c}/dB$ for a small amplitude of magnetic field drive. A magnetic field excitation $\delta B=\delta B_{\rm 0}\cos (2\pi  f_{\rm dr}^{\it B} t)$ of frequency $f_{\rm dr}^{\it B}$ is applied, with a small enough amplitude to remain in the linear regime.

At a fixed drive frequency $f_{\rm B}$, the amplitude of the voltage response $V_{\rm 0}^{\it B}(T)$ exhibits a peak in response at a temperature of about 4.7 K. This behavior is similar to the one observed in \mbox{Fig. \ref{fig:Vout_T}} in the WL device, and a signature of stochastic resonance. As in the WL case, the sharp rise is related to the exponential increase of the cutoff frequency $f_{\rm c}$ with the temperature. Above the peak, both $dV_{\rm av}/dI$ and $dI_{\rm c}/dB$ decrease with increasing temperature, leading to a sharper decrease in $dV_{\rm av}/dB(T)$ compared to the case of a WL device.

\begin{figure}[t!]
	\centering
	\includegraphics[width=0.9\columnwidth]{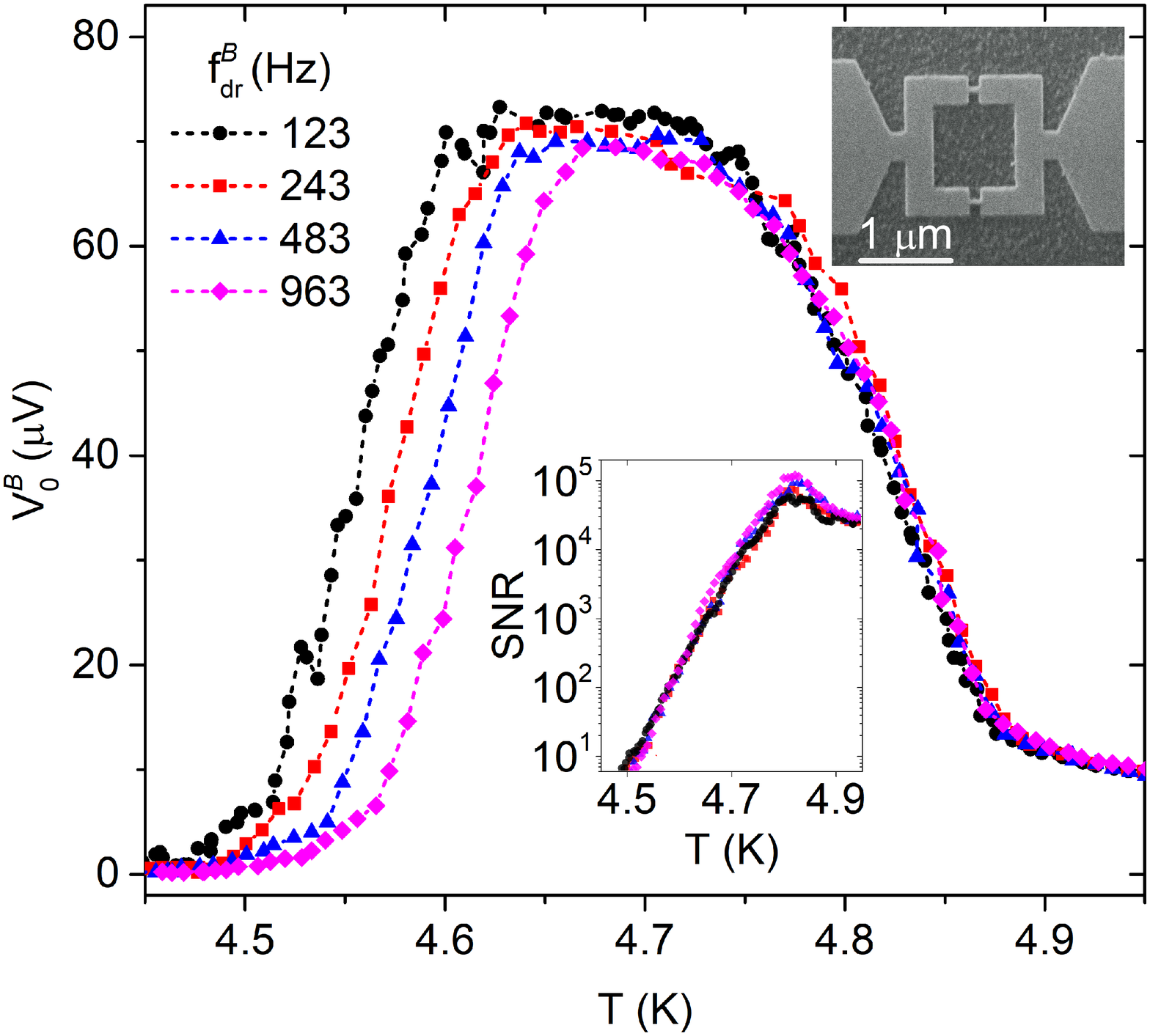}
	\caption{Amplitude of $\mu$-SQUID response $V_{\rm 0}^{\it B}$ to a periodic magnetic field drive of amplitude 6 $\mu$T as a function of bath temperature at few different drive frequencies $f_{\rm dr}^{\it B}$. The symbols are data and the dashed lines are guiding lines. The upper inset shows the scanning electron micrograph of the measured $\mu$-SQUID. The lower inset shows the experimentally obtained values of SNR as a function of temperature for each $f_{\rm dr}^{\it B}$.}
	\label{fig:Vout^B(fdr)}
\end{figure}

With increasing temperature, the voltage response keeps decreasing, see $V_{\rm 0}^{\it B}(T)$ in \mbox{Fig. \ref{fig:Vout^B(fdr)}}, while the noise power saturates to the white noise of the system. As a result, a peak in SNR is observed as shown in the inset of \mbox{Fig. \ref{fig:Vout^B(fdr)}}. The peak in SNR defines the best working region of the $\mu$-SQUID. At a temperature above 5 K, a non-hysteretic regime is observed, but with a lower SNR. We also discuss in the supplementary information \cite{suppl-info} that the flux sensitivity is enhanced in the bi-stable region, thanks to the large slope $dV_{\rm av}/dI$ of the $\mu$-SQUID IVC as compared to the dissipative {\textit{mono-stable}} region \cite{sourav-PRB-dyn}. The best flux resolution is found to be 10 $\mu\Phi_0/\sqrt{Hz}$, which includes the magnetic field noise as there is no magnetic shield in our system.

\section{Discussion and Conclusions}

In the frame of a stochastic resonance picture, the device response to a periodic drive remains large and nearly frequency independent up to a cutoff frequency $f_{\rm c}$. A high-$f_{\rm c}$ regime is thus preferred as it leads to higher signal-to-noise ratio and a larger measurement bandwidth. The exact optimum working parameters will also be determined by the amount of noise in the voltage signal from sources other than RTN. The crossover temperature $T_{\rm h}$ and hence the optimum working temperature can be tuned by changing the external shunt resistance and inductance \cite{sourav-ind-shunt}. A physical upper bound on the cutoff frequency $f_{\rm c}$ is set by the thermal relaxation time which sets the time-scale of deterministic dynamics and eventually limits the applicability of stochastic resonance at very high frequencies.

A possible drawback of the proposed scheme for nano-magnetism studies comes from the fluctuation in WL temperature (correlated with voltage RTN) which may influence the magnetic structure. This aspect is yet to be studied in detail as it is not clear if such temperature fluctuation will be confined to the electrons of the WL or transferred to the phonons, particularly for the high-$f_{\rm c}$ regime of interest here. In the former case, a poor electronic contact of the studied structure with the WL region may not be very difficult to achieve and enough to circumvent the issue.

In conclusion, other than illustrating the phenomenon of stochastic resonance quantitatively, our study unravels some important aspects about the thermally bistable regime of a WL. The direct evidence of an enhanced SNR within the bi-stable region opens new scenarios of noise induced improvement of the $\mu$-SQUID performance.

\section{Acknowledgments}

AKG acknowledges Sudeshna Sinha for a discussion on stochastic resonance. This work was supported by project 5804-2 from CEFIPRA, SERB-DST of the Government of India, ANR contract Optofluxonics \#17-CE30-0018 and LabEx LANEF (ANR-10-LABX-51-01) project UHV-NEQ. We are indebted to Thierry Crozes for help in the device fabrication at the Nanofab platform of N\'eel Institute.


\begin{thebibliography}:
\bibitem{Benzi_1981} R. Benzit, A. Sutera and A. Vulpiani, The mechanism of stochastic resonance, J. Phys. A: Math. Gen. {\bf 14}, L453 (1981).
\bibitem{wies-moss-persp} K. Wiesenfeld and F. Moss, Stochastic resonance and the benefits of noise: from ice ages to crayfish and SQUIDs, Nature {\bf 373}, 33 (1995).
\bibitem{fauve-heslot-ckt} S. Fauve and F. Heslot, Stochastic resonance in a bistable system, Phys. Lett. {\bf 97A}, 5 (1983).
\bibitem{vardi-double-dot} Y. Vardi, A. Guttman, and I. Bar-Joseph, Random Telegraph Signal in a Metallic Double-Dot System, Nano Lett. {\bf 14}, 2794 (2014).
\bibitem{wagner-nature} T. Wagner, P. Talkner, J. C. Bayer, E. P. Rugeramigabo, P. H\"{a}nggi, and R. J. Haug, Quantum stochastic resonance in an ac-driven single-electron quantum dot, Nature Phys. {\bf 15}, 330 (2019).
\bibitem{mcnamara-ring-laser} B. McNamara, K. Wiesenfeld, and R. Roy, Observation of Stochastic Resonance in a Ring Laser, Phys. Rev. Lett. {\bf 60}, 2626 (1988).
\bibitem{mantegna-tunn-diode} R. N. Mantegna and B. Spagnolo, Noise Enhanced Stability in an Unstable System, Phys. Rev. Lett. {\bf 76}, 563 (1996).
\bibitem{bradzey-nature} R. L. Badzey and P. Mohanty, Coherent signal amplification in bistable nanomechanical oscillators by stochastic resonance, Nature {\bf 437}, 995 (2005).
\bibitem{simon-optical-trap} A. Simon and A. Libchaber, Escape and Synchronization of a Brownian Particle, Phys. Rev. Lett. {\bf 68}, 3375 (1992).
\bibitem{hohman-SR-chem} W. Hohmann, J. M$\ddot{\rm u}$ller, and F. W. Schneider, Stochastic Resonance in Chemistry. 3. The Minimal-Bromate Reaction, J. Phys. Chem. {\bf 100}, 5388 (1996).
\bibitem{rf-SQUID-hibbs} A. D. Hibbs and A. L. Singsaas, E. W. Jacobs, A. R. Bulsara, J. J. Bekkedahl, and F. Moss, Stochastic resonance in a superconducting loop with a Josephson junction, J. Appl. Phys. {\bf 77}, 2582 (1995).
\bibitem{SQUID-rouse} R. Rouse, Siyuan Han, and J. E. Lukens, Flux amplification using stochastic superconducting quantum interference devices, Appl. Phys. Lett. {\bf 66}, 108 (1995).
\bibitem{RTN-PSD-machlup} S. Machlup, Noise in Semiconductors: Spectrum of a Two-Parameter Random Signal, J. Appl. Phys. {\bf 25}, 341 (1954).
\bibitem{Kramers} H. A. Kramers, Browninan motion in a field of force and the diffusion model of chemical reactions, Physica {\bf 7}, 284 (1940).
\bibitem{sudeshna-prl}K. Murali, S. Sinha, W. L. Ditto, A. R. Bulsara, Reliable Logic Circuit Elements that Exploit Nonlinearity in the Presence of a Noise-Floor, Phys. Rev. Lett. {\bf 102}, 104101 (2009).
\bibitem{likharev-rmp} K. K. Likharev, Superconducting weak links, Rev. Mod. Phys. {\bf 51}, 101 (1979).
\bibitem{rad-det} F. Marsili, F. Najafi, E. Dauler, F. Bellei, X. Hu, M. Csete, R. J. Molnar and K. K. Berggren, Single-photon detectors based on ultranarrow superconducting nanowires, Nano Lett. {\bf 11}, 2048 (2011); P. Virtanen, A. Ronzani, and F. Giazotto, Josephson Photodetectors via Temperature-to-Phase Conversion, Phys. Rev. Applied {\bf 9}, 054027 (2018).
\bibitem{SQUID-appl} W. Wernsdorfer, Adv. Chem. Phys. {\bf 118}, 99 (2001); D. Vasyukov, Y. Anahory, L. Embon, D. Halbertal, J. Cuppens, L. Neeman, A. Finkler, Y. Segev, Y. Myasoedov, M. L. Rappaport, M. E. Huber, and E. Zeldov, A scanning superconducting interference device with single electron spin resolution, Nature Nanotech. {\bf 8}, 639 (2013); T. Schwarz, R. W$\ddot{\rm o}$lbing, C. F. Reiche, B. M$\ddot{\rm u}$ller, M. J. Mart$\acute{\rm i}$nez-P$\acute{\rm e}$rez, T. M$\ddot{\rm u}$hl, B. B$\ddot{\rm u}$chner, R. Kleiner, and D. Koelle, Low-Noise YBa$_2$Cu$_3$O$_7$ Nano-SQUIDs for Performing Magnetization-Reversal Measurements on Magnetic Nanoparticles, Phys. Rev. Appl. {\bf 3}, 044011 (2015).
\bibitem{Nikhil-prl} N. Kumar, T. Fournier, H. Courtois, C. B. Winkelmann, and A. K. Gupta, Reversibility of superconducting Nb weak links driven by the proximity effect in a quantum interference device, Phys. Rev. Lett. {\bf 114}, 157003 (2015).
\bibitem{sourav-PRB-dyn} S. Biswas, C. B. Winkelmann, H. Courtois, and A. K. Gupta, Josephson coupling in the dissipative state of a thermally hysteretic micro-SQUID, Phys. Rev. B {\bf 98}, 174514 (2018).
\bibitem{DTM-JAP} A. K. Gupta, N. Kumar and S. Biswas, Temperature and phase dynamics in superconducting weak-link, J. Appl. Phys. {\bf 116}, 173901 (2014).
\bibitem{Scocpol} W. J. Skocpol, M. R. Beasley, and M. Tinkham, Self-heating hotspots in superconducting thin-film microbridges, J. Appl. Phys. {\bf 45}, 4054 (1974).
\bibitem{sourav-RTN} S. Biswas, N. Kumar, C. B. Winkelmann, H. Courtois, and A. K. Gupta, Random telegraphic voltage noise due to thermal bi-stability in a superconducting weak link,  AIP Conf. Proc. {\bf 1731}, 130001 (2016).
\bibitem{footnote-1} Note that on extremely short time scales ($\sim$ ps) well beyond experimental bandwidth and given by the Josephson frequency, the voltage in the dissipative state is itself time-dependent.
\bibitem{psd-suppl} For the spectral power we work in $f$ domain, rather than $\omega$ (=$2\pi f$), eliminating the need of $2\pi$ normalization factor. Further, the magnitude of spectral power does not depend on the sign of $f$ for the real measured signals. We use a convention in which we work with only positive $f$ with $S(f)$ values doubled to account for spectral-power of negative $f$. The average from $V(t)$ is also subtracted out before calculating DFT which makes $S(0)=0$. We summarize this in two relations: $S_{\rm V}(f)=2\int_{-\infty}^{\infty}A_{\rm V}(s)e^{-i 2\pi f s} ds$ and total spectral power, $\int_{0}^{\infty}S_{\rm V}(f)df=\sigma_{\rm V}^2$, with $A_{\rm V}(t)$ as the auto-correlation function of $V(t)$.
\bibitem{suppl-info} See Supplementary Information for mathematical details on two state model of stochastic resonance and some experimental details.
\bibitem{SR-rev-RMP} L. Gammaitoni, P. H\"{a}nggi, P. Jung and F. Marchesoni, Stochastic resonance, Rev. Mod. Phys. {\bf 70}, 223 (1998).
\bibitem{comment-DeltaU} From the Kramer's rate expression we see that for unequal $\omega_{\rm 1,2}$, the equality of $\tau_{\rm 1,2}$ at a particular $I_{\rm b}$, which controls $\Delta U_{\rm 1,2}$, will also depend on the noise intensity $D$. Both $\tau_{\rm i}$'s decrease with increasing $D$, following $\exp(-\Delta U_{\rm i}/D)$ and for, say, $\Delta U_{\rm 1}>\Delta U_{\rm 2}$, $\tau_{\rm 1}$ will decrease faster. In other words, for an increasing $D$ preserving the condition $\tau_{\rm 1}=\tau_{\rm 2}$ requires simultaneously raising $\Delta U_{\rm 1}$ and lowering $\Delta U_{\rm 2}$, see \mbox{Fig. \ref{fig:two-state}} inset, which is achieved by adjusting $I_{\rm b}$. Yet, $f_{\rm c}$ will increase with $D$, with a dependence in between $\exp(-\Delta U_1/D)$ and $\exp(-\Delta U_2/D)$.
\bibitem{tinkham-book} M. Tinkham, \textit{Introduction to Superconductivity}, 2nd ed. (Mc. Graw-Hill, New York, 1996).
\bibitem{PRL-Courtois-2008} H. Courtois, M. Meschke, J. T. Peltonen, and J. P. Pekola, Origin of Hysteresis in a Proximity Josephson Junction, Phys. Rev. Lett. {\bf 101}, 067002 (2008).
\bibitem{ben-jacob+valenti}E. Ben-Jacob, D. J. Bergman, and Z. Schuss, Thermal fluctuations and lifetime of the nonequilibrium steady state in a hysteretic Josephson junction, Phys. Rev. B {\bf 25}, 519 (1982); D. Valenti, C. Guarcello, and B. Spagnolo, Switching times in long-overlap Josephson junctions subject to thermal fluctuations and non-Gaussian noise sources, Phys. Rev. B {\bf 89}, 214510 (2014).
\bibitem{Shah-PRL} N. Shah, D. Pekker, and P. M. Goldbart, Phys. Rev. Lett. {\bf 101}, 207001 (2008).
\bibitem{gambling-foltyn} M. Foltyn and M. Zgirski, Gambling with Superconducting Fluctuations, Phys. Rev. Appl. {\bf 4}, 024002 (2015).
\bibitem{super-paramag} C. P. Bean and J. D. Livingston, Superparamagnetism, J. Appl. Phys. {\bf 30}, 120S (1959).
\bibitem{sourav-ind-shunt} S. Biswas, C. B. Winkelmann, H. Courtois, and A. K. Gupta, Elimination of thermal bistability in superconducting weak-links by an inductive shunt, Phys. Rev. B {\bf 101}, 024501 (2020).
\end{thebibliography}
\end{document}